# Effect of the forcing term in the pseudopotential lattice Boltzmann modeling of thermal flows

Q. Li[1] and K. H. Luo[2]

[1]Energy Technology Research Group, Faculty of Engineering and the Environment, University of Southampton, Southampton SO17 1BJ, United Kingdom

[2]Department of Mechanical Engineering, University College London, Torrington Place, London WC1E 7JE, United Kingdom

The pseudopotential lattice Boltzmann (LB) model is a popular model in the LB community for simulating multiphase flows. Recently, several thermal LB models, which are based on the pseudopotential LB model and constructed within the framework of the double-distribution-function LB method, were proposed to simulate thermal multiphase flows [G. Házi and A. Márkus, Phys. Rev. E **77**, 026305 (2008); L. Biferale *et al.*, Phys. Rev. Lett. **108**, 104502 (2012); S. Gong and P. Cheng, Int. J. Heat Mass Transfer **55**, 4923 (2012)]. The objective of the present paper is to show that the effect of the forcing term on the temperature equation must be eliminated in the pseudopotential LB modeling of thermal flows. First, the effect of the forcing term on the temperature equation is shown via the Chapman-Enskog analysis. For comparison, alternative treatments that are free from the forcing-term effect are provided. Subsequently, numerical investigations are performed for two benchmark tests. The numerical results clearly show that the existence of the forcing-term effect will lead to significant numerical errors in the pseudopotential LB modeling of thermal flows.

## I. Introduction

In the past two decades, the lattice Boltzmann (LB) method, which historically originates from the lattice gas automata [1], has been developed into an efficient mesoscopic numerical approach for simulating fluid flow and heat transfer [2-5]. Compared with the traditional numerical methods based on the discretization of the Navier-Stokes equations, the LB method has some distinctive advantages, such as the simple form of the governing equations, the easiness of programming, the avoidance of nonlinear convective terms, and the inherent parallelizability on multiple processors.

Owing to its distinctive advantages, the LB method is found to be promising for simulating multiphase flows. Many multiphase LB models have been developed from a variety of points of view [2, 4, 5]. Among these models, the pseudopotential LB model proposed by Shan and Chen [6] has attracted much attention because of its simplicity and the kinetic nature that the phase segregation can emerge naturally in the pseudopotential LB model as a result of particle interactions, without tracking or capturing the interfaces between different phases [7].

In recent years, several thermal LB models [8-13], which are based on the pseudopotential LB model, were proposed by Házi *et al*. [8-10], Biferale *et al*. [11], and Cheng *et al*. [12, 13], respectively, for simulating thermal multiphase flows. Most recently, by defining more complete and reasonable macroscopic source terms in the energy equation, Kamali *et al*. [14] have also proposed a thermal LB model based on the pseudopotential LB model. From Refs. [8-14] it can be found that these models share the feature that they are all constructed within the framework of the double-distribution-function LB method: a density distribution function is used to solve the density and velocity fields, while the temperature field is solved by another set of distribution function.

Actually, for thermal LB equations, the forcing term of the system will introduce an additional term into the macroscopic temperature equation. Such an additional term exists in many thermal LB models based on the temperature (or internal energy) distribution function [15-19] and also in the above-mentioned simulations of thermal multiphase flows [8-14] on the basis of the pseudopotential LB model, although this term was not shown in these studies. In most cases, the errors caused by the additional term are very small. However, in the present paper we shall show that the additional term, which represents the effect of the forcing term on the temperature equation, must be eliminated in the pseudopotential LB modeling of thermal flows because it will lead to significant numerical errors.

The rest of the present paper is organized as follows. In Sec. II, the pseudopotential LB model is briefly introduced. The effect of the forcing term on the temperature equation will be revealed through the Chapman-Enskog analysis. For comparison, alternative treatments free from the forcing-term effect are also provided. Numerical analyses will be conducted in Sec. III. and finally a brief conclusion will be made in Sec. IV.

## II. Theoretical analyses

### A. The pseudopotential LB model

The LB equation with the Bhatnagar-Gross-Krook collision operator [20] can be written as follows:

$$f_i(\mathbf{x}+\mathbf{e}_i\delta_t, t+\delta_t) - f_i(\mathbf{x},t) = -\frac{1}{\tau_f}\left[f_i(\mathbf{x},t) - f_i^{eq}(\mathbf{x},t)\right] + \delta_t F_i(\mathbf{x},t), \quad (1)$$

where $f_i$ is the density distribution function, $f_i^{eq}$ is its equilibrium distribution given by $f_i^{eq} = \rho\omega_i\left[1+(\mathbf{e}_i\cdot\mathbf{u})/c_s^2 + 0.5(\mathbf{e}_i\cdot\mathbf{u})^2/c_s^4 - 0.5|\mathbf{u}|^2/c_s^2\right]$, $\tau_f$ is the corresponding relaxation time, $t$ is the time, $\mathbf{x}$ is the spatial position, $\mathbf{e}_i$ is the discrete velocity along the $i$ direction, $\delta_t$ is the time step, and $F_i$ is the forcing term, which is used to incorporate a force $\mathbf{F}$ into the system [21]. In the pseudopotential LB model, the pseudopotential force $\mathbf{F}$ is often given by [22, 23]

$$\mathbf{F} = -G\psi(\mathbf{x}) \sum_{i=1}^{8} w\left(|\mathbf{e}_i|^2\right) \psi(\mathbf{x}+\mathbf{e}_i) \mathbf{e}_i, \tag{2}$$

where $\psi$ is the pseudopotential and $w\left(|\mathbf{e}_i|^2\right)$ are the weights. For the nearest-neighbor interactions on the D2Q9 lattice, the weights are $w(1)=1/3$ and $w(2)=1/12$. Using the Taylor series expansion, it can be found that Eq. (2) gives

$$\mathbf{F} = -Gc^2 \left[ \frac{1}{2} \nabla \psi^2 + \frac{1}{6} c^2 \psi \nabla\left(\nabla^2 \psi\right) + \cdots \right], \tag{3}$$

where $c=1$ is the lattice constant. The high-order term in Eq. (3) yields the surface tension for multiphase fluids. To obtain a desired equation of state, the pseudopotential $\psi$ is usually chosen as $\psi(\rho) = \sqrt{2(p_{\text{EOS}} - p)/Gc^2}$ [24, 25], where $p_{\text{EOS}}$ is the desired equation of state, while $p = \rho c_s^2$ ($c_s = c/\sqrt{3}$) is the equation of state in the standard LB method. In some studies [10, 12, 13], a mixed scheme is adopted for the pseudopotential force

$$\mathbf{F} = -G \left[ \beta \psi(\mathbf{x}) \sum_{i=1}^{8} w\left(|\mathbf{e}_i|^2\right) \psi(\mathbf{x}+\mathbf{e}_i) \mathbf{e}_i + \frac{1-\beta}{2} \sum_{i=1}^{8} w\left(|\mathbf{e}_i|^2\right) \psi^2(\mathbf{x}+\mathbf{e}_i) \mathbf{e}_i \right]. \tag{4}$$

It can be found that both Eq. (2) and Eq. (4) satisfy $\mathbf{F} \approx -Gc^2 \nabla \psi^2/2 = -\nabla\left(p_{\text{EOS}} - \rho c_s^2\right)$.

### *B. The effect of the forcing term on the temperature equation*

As previously mentioned, several thermal LB models were recently proposed based on the pseudopotential LB model for simulating thermal multiphase flows. In these models, the temperature field is solved by another set of distribution function $g_\alpha$. The target temperature equation can be written as follows [8-10]:

$$\partial_t(\rho c_v T) + \nabla \cdot (\rho c_v T \mathbf{u}) = \nabla \cdot (\lambda \nabla T) + \phi \nabla \cdot \mathbf{u}, \tag{5}$$

where $\lambda$ is the thermal conductivity, $c_v$ is the specific heat at constant volume, and $\phi = -T\left(\partial p_{\text{EOS}}/\partial T\right)_\rho$ [8]. In these models, the term $\phi \nabla \cdot \mathbf{u}$ is realized by incorporating a source term into the thermal LB equation. For simplicity, the term $\phi \nabla \cdot \mathbf{u}$ is omitted in the present study and such a choice will not affect our analyses. The temperature equation is then given by

$$\partial_t (\rho c_v T) + \nabla \cdot (\rho c_v T \mathbf{u}) = \nabla \cdot (\lambda \nabla T). \tag{6}$$

In the LB community, thermal LB equations for solving Eq. (6) can be found everywhere [18, 19]:

$$g_i(\mathbf{x} + \mathbf{e}_i \delta_t, t + \delta_t) - g_i(\mathbf{x}, t) = -\frac{1}{\tau_g} \left[ g_i(\mathbf{x}, t) - g_i^{eq}(\mathbf{x}, t) \right], \tag{7}$$

where the equilibrium distribution function $g_i^{eq}$ can be defined as $g_i^{eq} = c_v T f_i^{eq}$.

For thermal LB equations, the forcing term in Eq. (1) will introduce an additional term into the macroscopic temperature equation. To display this forcing-term effect clearly, the Chapman-Enskog analysis of Eq. (7) is given here for general readers. Through the Taylor series expansion, Eq. (7) will become

$$\delta_t (\partial_t + \mathbf{e}_i \cdot \nabla) g_i + \frac{\delta_t^2}{2} (\partial_t + \mathbf{e}_i \cdot \nabla)^2 g_i + \cdots = -\frac{1}{\tau_g} (g_i - g_i^{eq}). \tag{8}$$

Using the following multi-scale expansions

$$\partial_t = \partial_{t0} + \delta_t \partial_{t1}, \quad g_i = g_i^{eq} + \delta_t g_i^{(1)} + \delta_t^2 g_i^{(2)}, \tag{9}$$

Eq. (8) can be rewritten in the consecutive orders of $\delta_t$ as follows:

$$O(\delta_t): \quad (\partial_{t0} + \mathbf{e}_i \cdot \nabla) g_i^{eq} = -\frac{1}{\tau_g} g_i^{(1)}, \tag{10}$$

$$O(\delta_t^2): \quad \partial_{t1} g_i^{eq} + (\partial_{t0} + \mathbf{e}_i \cdot \nabla) g_i^{(1)} + \frac{1}{2} (\partial_{t0} + \mathbf{e}_i \cdot \nabla)^2 g_i^{eq} = -\frac{1}{\tau_g} g_i^{(2)}. \tag{11}$$

According to Eq. (10), we can rewrite Eq. (11) as

$$\partial_{t1} g_i^{eq} + (\partial_{t0} + \mathbf{e}_i \cdot \nabla) \left( 1 - \frac{1}{2\tau_g} \right) g_i^{(1)} = -\frac{1}{\tau_g} g_i^{(2)}. \tag{12}$$

Taking the summations of Eq. (10) and Eq. (12), the following equations can be obtained, respectively:

$$\partial_{t0} (\rho c_v T) + \nabla \cdot (\rho c_v T \mathbf{u}) = 0, \tag{13}$$

$$\partial_{t1} (\rho c_v T) + \nabla \cdot \left( 1 - \frac{1}{2\tau_g} \right) \left( \sum_i \mathbf{e}_i g_i^{(1)} \right) = 0. \tag{14}$$

According to Eq. (10), we can obtain

$$\sum_i \mathbf{e}_i g_i^{(1)} = -\tau_g \left[ \partial_{t0} \left( \sum_i \mathbf{e}_i g_i^{eq} \right) + \nabla \cdot \left( \sum_i \mathbf{e}_i \mathbf{e}_i g_i^{eq} \right) \right]. \tag{15}$$

Note that $g_i^{eq} = c_v T f_i^{eq}$, hence we have

$$\partial_{t0}\left(\sum_i \mathbf{e}_i g_i^{eq}\right) = \partial_{t0}(\rho c_v T \mathbf{u}) \equiv \mathbf{u}\partial_{t0}(\rho c_v T) + \rho c_v T \partial_{t0}\mathbf{u}, \tag{16}$$

$$\nabla \cdot \left(\sum_i \mathbf{e}_i \mathbf{e}_i g_i^{eq}\right) = \nabla \cdot (\rho c_v T \mathbf{u}\mathbf{u}) + c_v p \nabla T + c_v T \nabla p, \tag{17}$$

where $p = \rho c_s^2$. In Eq. (16), the term $\partial_{t0}(\rho c_v T)$ can be obtained from Eq. (13), while the term $\rho \partial_{t0}\mathbf{u}$ should be evaluated as follows:

$$\rho \partial_{t0}\mathbf{u} = \partial_{t0}(\rho \mathbf{u}) - \mathbf{u}\partial_{t0}\rho. \tag{18}$$

Both $\partial_{t0}(\rho \mathbf{u})$ and $\partial_{t0}\rho$ are related to the Chapman-Enskog analysis of the LB equation for the density distribution function, namely Eq. (1), which is an usual procedure in the LB community and the following results can be readily obtained:

$$\partial_{t0}\rho + \nabla \cdot (\rho \mathbf{u}) = 0, \quad \partial_{t0}(\rho \mathbf{u}) + \nabla \cdot (\rho \mathbf{u}\mathbf{u}) = -\nabla p + \mathbf{F}. \tag{19}$$

Using Eq. (19), we can obtain

$$\rho \partial_{t0}\mathbf{u} = -\rho \mathbf{u} \cdot \nabla \mathbf{u} - \nabla p + \mathbf{F}. \tag{20}$$

According to Eqs. (13) and (20), Eq. (16) can be rewritten as

$$\partial_{t0}\left(\sum_i \mathbf{e}_i g_i^{eq}\right) = -\mathbf{u}\nabla \cdot (\rho c_v T \mathbf{u}) - \rho c_v T \mathbf{u} \cdot \nabla \mathbf{u} - c_v T \nabla p + c_v T \mathbf{F}. \tag{21}$$

Substituting Eqs. (21) and (17) into Eq. (15), we have

$$\sum_i \mathbf{e}_i g_i^{(1)} = -\tau_g \left(c_v p \nabla T + c_v T \mathbf{F}\right). \tag{22}$$

Combining Eq. (13) with Eq. (14) and using Eq. (22), we can obtain

$$\partial_t (\rho c_v T) + \nabla \cdot (\rho c_v T \mathbf{u}) = \nabla \cdot (\lambda \nabla T + \vartheta T \mathbf{F}), \tag{23}$$

where $\lambda = (\tau_g - 0.5)c_v p$ is thermal conductivity and $\vartheta = \lambda/p$. Obviously, compared with Eq. (6), Eq. (23) contains an unwanted term $\nabla \cdot (\vartheta T \mathbf{F})$, which is just *the effect of the forcing term* on the temperature equation.

It is clear that the thermal LB equation (7) solves Eq. (23) rather than Eq. (6). However, from the literature, it can be found that using Eq. (7) or its variations to mimic Eq. (6) has been widely practiced in the LB community. In most cases, the numerical errors caused by the additional term in Eq. (23) are very small. Unfortunately, this is not true for the pseudopotential LB model. The problem arises from the fact that the force $\mathbf{F}$ in the pseudopotential LB model may enable the term $\vartheta T\mathbf{F}$ to be comparable with the heat flux term $\lambda \nabla T$.

To numerically quantify the forcing-term effect, two treatments free from this effect are provided for comparison. A simple treatment is adding a correction term into the thermal LB equation so as to eliminate the unwanted term in Eq. (23). The corrected thermal LB equation is then given by

$$g_i(\mathbf{x}+\mathbf{e}_i\delta_t, t+\delta_t) - g_i(\mathbf{x},t) = -\frac{1}{\tau_g}\left[g_i(\mathbf{x},t) - g_i^{eq}(\mathbf{x},t)\right] + \delta_t C_i(\mathbf{x},t), \qquad (24)$$

where $C_i$ is the correction term

$$C_i = \left(1 - \frac{1}{2\tau_g}\right)\omega_i c_v T \frac{(\mathbf{e}_i \cdot \mathbf{F})}{c_s^2}. \qquad (25)$$

It can be found that $\sum_i C_i = 0$, $\sum_i \mathbf{e}_i C_i = (1-0.5/\tau_g)c_v T\mathbf{F}$, and $\sum_i \mathbf{e}_i\mathbf{e}_i C_i = 0$. The feature that $\sum_i C_i = 0$ distinguishes the correction term $C_i$ from the usual source terms in the thermal LB equation, which are employed to recover the macroscopic source terms in the target temperature equation [8-14]. The Chapman-Enskog analysis of Eq. (24) is given in the Appendix, which shows that the required temperature equation can be correctly recovered. Actually, in the literature the first author of the present paper and his coworkers [26] have briefly mentioned that an additional term is needed for thermal LB equations in the presence of a body force.

Furthermore, another treatment is also considered: using the finite-difference method to solve Eq. (6), which can be rewritten as

$$\partial_t T = -\mathbf{u}\cdot \nabla T + \frac{1}{\rho c_v}\nabla \cdot (\lambda \nabla T) \equiv K(T). \qquad (26)$$

The second-order Runge-Kutta scheme is adopted for time discretization:

$$T^{n+1} = T^n + \frac{\delta_t}{2}(h_1 + h_2), \quad h_1 = K(T^n), \quad h_2 = K\left(T^n + \frac{\delta_t}{2}h_1\right). \tag{27}$$

The isotropic central schemes are employed to evaluate the first-order derivative and the Laplacian [27]. It is expected that the forcing-term effect on the temperature equation can be quantified by comparing the results obtained by Eq. (7) with the results of Eqs. (24) and (26).

## III. Numerical results

In this section, numerical simulations are carried out to investigate the forcing-term effect in the pseudopotential LB modeling of thermal flows. Due to the complexity of thermal multiphase flows with the non-ideal equations of state such as $p_{\text{EOS}} = \rho RT/(1-b\rho) - a\rho^2$, available analytical solutions and benchmark tests are very rare. This is the reason why few quantitative validations were given in Refs. [8-13]. To achieve quantitative comparisons, in the present study the ideal equation of state $p_{\text{EOS}} = \rho RT$ is utilized, which corresponds to single-phase fluids. To minimize the influence of the higher-order terms (which yield the surface tension for multiphase fluids but give errors for single-phase fluids) in the force, we use $\beta = 0$ in Eq. (4). Then $\mathbf{F} = -0.5G \sum_{i=1}^{8} w(|\mathbf{e}_i|^2)\psi^2(\mathbf{x}+\mathbf{e}_i)\mathbf{e}_i$, where $\psi^2(\rho) = 2(p_{\text{EOS}} - \rho c_s^2)/Gc^2$. For $\psi^2(\rho)$, the parameter $G$ can be taken as $G = 1$.

### A. Planar flow between parallel plates

First, we consider a two-dimensional planar flow between parallel plates at rest. Uniform velocity and temperature profiles, $U_{\text{in}}$ and $T_{\text{in}}$, are applied at the inlet, while the hydrodynamically and thermally fully developed condition is imposed at the outlet. Two different cases of thermal boundary conditions are considered at the plates. *Case A*: the upper and lower plates are kept at the uniform temperature $T_w$; and *Case B*: the upper plate is kept at the temperature $T_w$ while the lower plate is adiabatic ($q_w = 0$). For these two cases of thermal boundary conditions, the corresponding analytical

Nusselt numbers in the thermally fully developed region are $\text{Nu} = 7.54$ and $\text{Nu} = 4.86$ [28], respectively.

In simulations, a $N_x \times N_y = 500 \times 60$ lattice system is adopted. The dynamic viscosity $\mu = \rho c_s^2 (\tau_f - 0.5)$ is set to 0.1 and the Prandtl number is fixed at $\text{Pr} = 0.71$. The parameter $T_{\text{in}}$, $T_{\text{w}}$, $U_{\text{in}}$, $c$, and $\delta_t$ are chosen as follows (in lattice units): $T_{\text{in}} = 1.03$, $T_{\text{w}} = 0.97$, $U_{\text{in}} = 0.05$, $c = 1$, and $\delta_t = 1$. The characteristic temperature $T_c$ is taken as $T_c = (T_{\text{in}} + T_{\text{w}})/2$ and then $R$ is determined via $c = \sqrt{3RT_c} = 1$ [16]. It can be seen that the temperature variation is very small: $(1 \pm 0.03) T_c$. Under such a condition, the flow is near the incompressible limit. The obtained local Nusselt numbers along the flow direction are plotted in Fig. 1. The local Nusselt number is defined as $\text{Nu}(x) = D_h q_w / [\lambda (T_w - T_b)]$, where $D_h$ is the hydraulic diameter of the channel, $q_w = \lambda (\partial T / \partial y)_w$ is the local heat flux at the wall, $\lambda$ is the thermal conductivity, and $T_b = \int_0^H \rho u_x T \, dy / \int_0^H \rho u_x \, dy$ is the local bulk temperature.

From Fig. 1 we can see that in both cases there are no apparent differences between the results obtained by the corrected thermal LB Eq. (24) and the finite-difference solution of Eq. (26), while the results given by the thermal LB Eq. (7) significantly deviate from the results of Eqs. (24) and (26). Quantitatively, the predicted Nusselt numbers in thermally fully developed region given by Eqs. (24) and (26) are compared with the analytical solutions in Table I, from which good agreement can be observed. On the contrary, in the region near the outlet the Nusselt numbers obtained by Eq. (7) are around $\text{Nu} = 13$ and 12.5 for Cases A and B, respectively, which are much larger than the corresponding analytical results.

### *B. Natural convection in a square cavity*

Now we consider another test: the natural convection in a two-dimensional square cavity. In this problem, the sidewalls of the cavity are maintained at constant but different temperatures, whereas the bottom and top walls are adiabatic. The natural convection can be characterized by the Prandtl number and the Rayleigh number, which is defined as [29]

$$\text{Ra} = \frac{g\beta\rho_0^2(T_h - T_l)L^3 \text{Pr}}{\mu^2}, \quad (28)$$

where $g$ is the gravity acceleration, $T_h$ and $T_l$ are the temperatures of the left and right walls, respectively, $L$ is the distance between the walls, and $\beta = 1/T_c$ is the thermal expansion coefficient, where $T_c = (T_h + T_l)/2$ is the characteristic temperature.

Since the ideal equation of state $p_{\text{EOS}} = \rho RT$ is employed in the pseudopotential, the Boussinesq assumption is not needed and the buoyancy force $\mathbf{G} = (0, -\rho g)$ can be directly added to the forcing term. The temperatures $T_h$ and $T_l$ are chosen as $T_h = 1.03$ and $T_l = 0.97$. The dynamic viscosity $\mu$ is set to $\mu = 0.1$ and the Prandtl number is fixed at $\text{Pr} = 0.71$. Two different Rayleigh numbers are considered: $\text{Ra} = 10^3$ and $10^4$. The corresponding lattice systems are $N_x \times N_y = 100 \times 100$ and $150 \times 150$, respectively.

The isotherms given by the three different treatments are illustrated in Fig. 2. From the figure we can see that the results obtained by the corrected thermal LB Eq. (24) and the finite-difference solution of Eq. (26) are nearly the same, while the results given by Eq. (7) are obviously different. To be specific, from Fig. 2(a) we can find that in the results of Eq. (7) the temperature gradients ($\partial_x T$) near the left and right walls are much larger than those in Figs. 2(b) and 2(c). Meanwhile, due to the coupling between the velocity and the temperature fields, the streamlines given by Eq. (7) also significantly deviate from the results of Eqs. (24) and (26), which can be seen in Fig. 3.

In fact, according to Eq. (23) we have monitored the coefficient $\lambda_{\text{eff},x} = (\lambda \partial_x T + \vartheta T F_x)/\partial_x T$ and found that, near the left and right walls, $\lambda_{\text{eff},x}$ is very small as compared with $\lambda$. In other words, the forcing-term effect will cause the modeled Rayleigh numbers near the left and right walls to be much higher than the defined Rayleigh number. This is the reason why in Fig. 2(a) the isotherms near the left and right walls are very dense. Quantitatively, the average Nusselt number at the hot wall is computed. The results obtained by the three different treatments are listed in Table II together with the benchmark solution in Ref. [30]. As can be seen in Table II, the results of Eqs. (24) and (26) are in good agreement with the data reported by Barakos *et al*. [30], while the Nusselt numbers given by Eq. (7) are apparently inaccurate. Specifically, the relative error at $\text{Ra} = 10^4$ is larger than 250%.

### IV. Conclusions

In summary, we have investigated the effect of the forcing term on the temperature equation in the pseudopotential LB modeling of thermal flows. First, theoretical analyses have been conducted to reveal the forcing-term effect on the temperature equation. It is shown that, due to the forcing-term effect, an unwanted term $\nabla \cdot (\vartheta T \mathbf{F})$ exists in the macroscopic temperature equation. Numerical analyses have been carried out for two benchmark tests: thermally fully developed flows between parallel plates and the natural convection in a square cavity. The numerical results clearly show that the existence of the forcing-term effect on the temperature equation will lead to significant numerical errors.

On the basis of the numerical results, we can conclude that the forcing-term effect on the temperature equation must be eliminated in the pseudopotential LB modeling of thermal flows. It has been shown that, within the double-distribution-function LB framework, the forcing-term effect can be eliminated by adding a correction term into the thermal LB equation. Meanwhile, the forcing-term effect can also be avoided by using traditional numerical methods such as the finite-difference method to solve the temperature field, which falls into the hybrid thermal LB framework. Furthermore, it is worth mentioning that the multispeed high-order LB approach [31-33], which is another approach for constructing thermal LB models [34-36], also does not suffer from the mentioned problem when a correct forcing term is employed.

**APPENDIX: THE CHAPMAN-ENSKOG ANALYSIS OF EQ. (24)**

The Chapman-Enskog analysis of Eq. (24) is similar to that of Eq. (7). Firstly, through the Taylor series expansion, Eq. (24) will yield

$$\delta_t \left( \partial_t + \mathbf{e}_i \cdot \nabla \right) g_i + \frac{\delta_t^2}{2} \left( \partial_t + \mathbf{e}_i \cdot \nabla \right)^2 g_i + \cdots = -\frac{1}{\tau_g} \left( g_i - g_i^{eq} \right) + \delta_t C_i, \quad (A1)$$

Using Eq. (9), we can rewrite Eq. (A1) in the consecutive orders of $\delta_t$ as follows:

$$O(\delta_t): \left( \partial_{t0} + \mathbf{e}_i \cdot \nabla \right) g_i^{eq} = -\frac{1}{\tau_g} g_i^{(1)} + C_i, \quad (A2)$$

$$O(\delta_t^2): \partial_{t1} g_i^{eq} + \left( \partial_{t0} + \mathbf{e}_i \cdot \nabla \right) g_i^{(1)} + \frac{1}{2} \left( \partial_{t0} + \mathbf{e}_i \cdot \nabla \right)^2 g_i^{eq} = -\frac{1}{\tau_g} g_i^{(2)}. \quad (A3)$$

With the help of Eq. (A2), Eq. (A3) can be rewritten as

$$\partial_{t1} g_i^{eq} + \left( \partial_{t0} + \mathbf{e}_i \cdot \nabla \right) \left( 1 - \frac{1}{2\tau_g} \right) g_i^{(1)} + \frac{1}{2} \left( \partial_{t0} + \mathbf{e}_i \cdot \nabla \right) C_i = -\frac{1}{\tau_g} g_i^{(2)}. \quad (A4)$$

Taking the summations of Eq. (A2) and Eq. (A4) leads to, respectively

$$\partial_{t0} \left( \rho c_v T \right) + \nabla \cdot \left( \rho c_v T \mathbf{u} \right) = 0, \quad (A5)$$

$$\partial_{t1} \left( \rho c_v T \right) + \nabla \cdot \left[ \left( 1 - \frac{1}{2\tau_g} \right) \sum_i \mathbf{e}_i g_i^{(1)} + \frac{1}{2} \sum_i \mathbf{e}_i C_i \right] = 0. \quad (A6)$$

From Eq. (A2), the following equation can be obtained:

$$\sum_i \mathbf{e}_i g_i^{(1)} = -\tau_g \left[ \partial_{t0} \left( \sum_i \mathbf{e}_i g_i^{(0)} \right) + \nabla \cdot \left( \sum_i \mathbf{e}_i \mathbf{e}_i g_i^{(0)} \right) - \sum_i \mathbf{e}_i C_i \right]. \quad (A7)$$

According to Eqs. (21) and (17), we can obtain

$$\partial_{t0}\left(\sum_i \mathbf{e}_i g_i^{(0)}\right)+\nabla\cdot\left(\sum_i \mathbf{e}_i \mathbf{e}_i g_i^{(0)}\right)= pc_v\nabla T + c_v T\mathbf{F}. \tag{A8}$$

From Eqs. (A7) and (A8), we have

$$\left(1-\frac{1}{2\tau_g}\right)\sum_i \mathbf{e}_i g_i^{(1)}+\frac{1}{2}\sum_i \mathbf{e}_i C_i = -\left(\tau_g-\frac{1}{2}\right)\left(pc_v\nabla T+c_v T\mathbf{F}\right)+\tau_g \sum_i \mathbf{e}_i C_i . \tag{A9}$$

Substituting Eq. (A9) into Eq. (A6) and noting that $\sum_i \mathbf{e}_i C_i = \left(1-0.5/\tau_g\right)c_v T\mathbf{F}$, we can obtain

$$\partial_{t1}(\rho c_v T) = \nabla\cdot(\lambda\nabla T). \tag{A10}$$

With Eqs. (A5) and (A10), the following macroscopic temperature can be recovered:

$$\partial_t(\rho c_v T)+\nabla\cdot(\rho c_v \mathbf{u} T) = \nabla\cdot(\lambda\nabla T). \tag{A11}$$

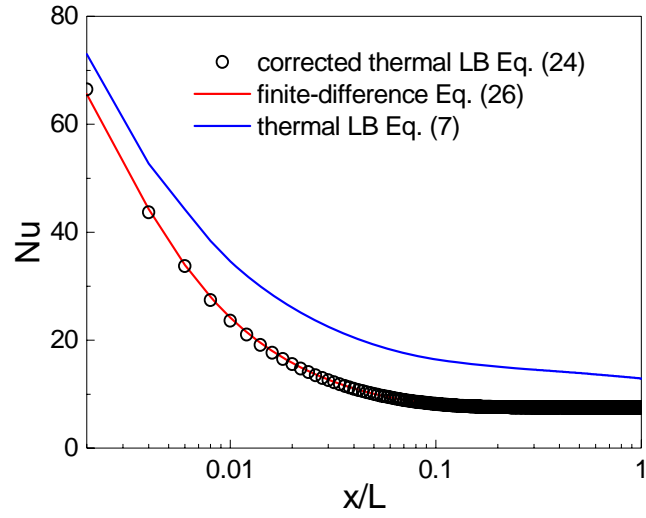

(a) Case A

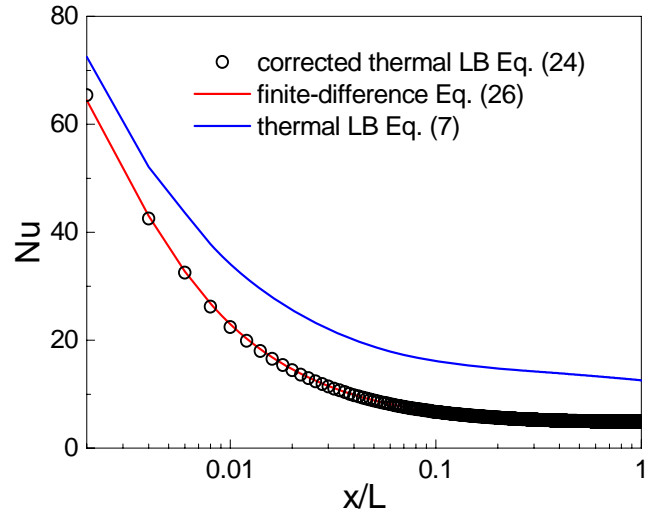

(b) Case B

FIG. 1. The local Nusselt number distribution along the flow direction for two different cases of thermal boundary conditions. Case A: the upper and lower plates are kept at the temperature $T_w$; and Case B: the upper plate is kept at the temperature $T_w$ while the lower plate is adiabatic.

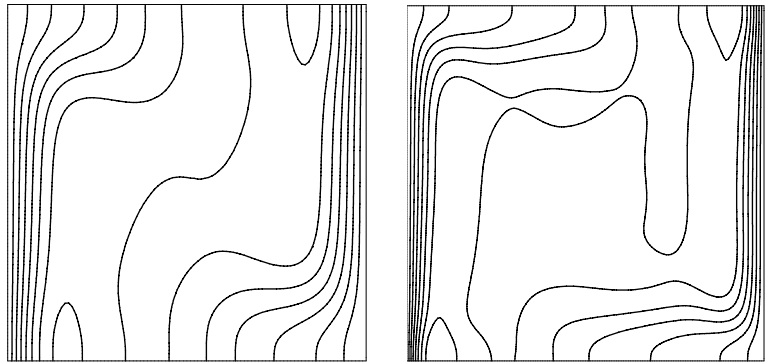
(a) thermal LB Eq. (7)

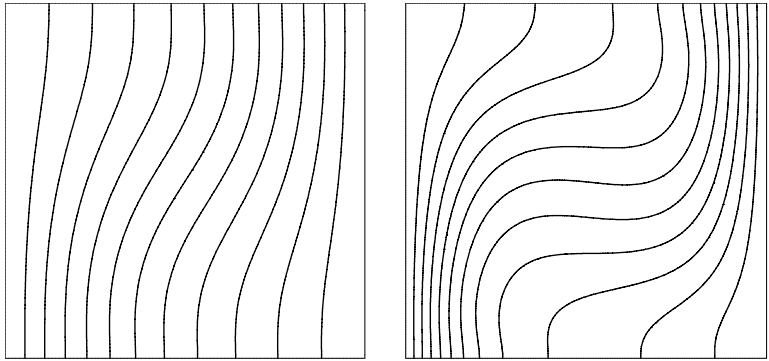
(b) corrected thermal LB Eq. (24)

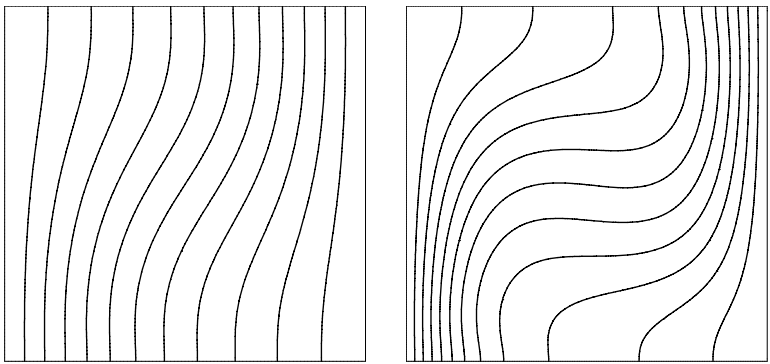
(c) finite-difference Eq. (26)

FIG. 2. Isotherms obtained by different treatments at $\mathrm{Ra}=10^3$ (left) and $\mathrm{Ra}=10^4$ (right).

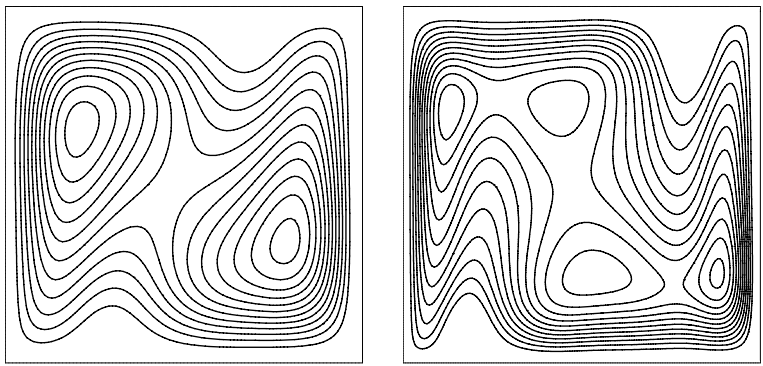

(a) thermal LB Eq. (7)

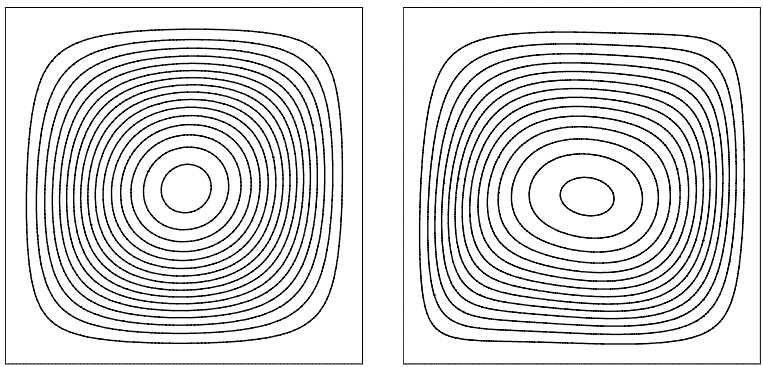

(b) corrected thermal LB Eq. (24)

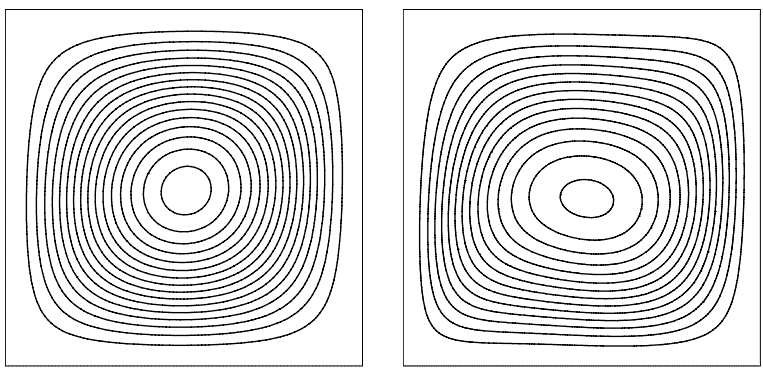

(c) finite-difference Eq. (26)

FIG. 3. Streamlines obtained by different treatments at $Ra = 10^3$ (left) and $Ra = 10^4$ (right).

Table I. Comparison of the Nusselt number in the thermally fully developed region.

| Case | corrected thermal LB Eq. (24) | finite-difference Eq. (26) | analytical [28] |
|---|---|---|---|
| A | 7.51 | 7.51 | 7.54 |
| B | 4.91 | 4.91 | 4.86 |

Table II. Natural convection in a square cavity: comparison of the average Nusselt number.

| Ra | thermal LB Eq. (7) | corrected thermal LB Eq. (24) | finite-difference Eq. (26) | Barakos *et al.* [30] |
|---|---|---|---|---|
| $10^3$ | 4.884 | 1.118 | 1.118 | 1.114 |
| $10^4$ | 8.431 | 2.246 | 2.242 | 2.245 |